\documentclass[aps,prl,twocolumn,floatfix,superscriptaddress,showpacs,amsmath,amssymb]{revtex4}
\usepackage{amsmath}
\usepackage{graphicx}
\usepackage{amsfonts}
\makeatletter
\def\captionof#1#2{{\def\@captype{#1}#2}}
\makeatother
\newcommand{\rmc}{{\rm c}}
\newcommand{\rme}{{\rm e}}
\newcommand{\rmd}{{\rm d}}
\newcommand{\rmi}{{\rm i}}
\newcommand{\rmR}{{\rm R}}
\newcommand{\rmq}{{\rm q}}

\begin{document}

\title{Steady-state entanglement of spatially separated qubits \textit{via} quantum bath engineering}
\author{Camille Aron}
\affiliation{Department of Electrical Engineering, Princeton University, Princeton, NJ 08544, USA}
\author{Manas Kulkarni}
\affiliation{Department of Physics, New York City College of
Technology, The City University of New York, Brooklyn, NY 11201, USA}
\affiliation{Department of Electrical Engineering, Princeton University, Princeton, NJ 08544, USA}
\author{Hakan E. T\"{u}reci}
\affiliation{Department of Electrical Engineering, Princeton University, Princeton, NJ 08544, USA}

\begin{abstract}
We propose a scheme for driving a dimer of spatially separated qubits into a maximally entangled non-equilibrium steady state. A photon-mediated retarded interaction between the qubits is realized by coupling them to two tunnel-coupled leaky cavities where each cavity is driven by a coherent microwave tone. The proposed cooling mechanism relies on striking the right balance between the unitary and driven-dissipative dynamics of the qubit subsystem. 
We map the dimer to an effective transverse-field $XY$ model coupled to a non-equilibrium bath that can be suitably engineered through the choice of drive frequencies and amplitudes. We show that both singlet and triplet states can be obtained with remarkable fidelities.
 The proposed protocol can be implemented with a superconducting circuit architecture that was recently experimentally realized and paves the way to achieving large-scale entangled systems that are arbitrarily long lived.
\end{abstract}

\maketitle

\paragraph{Introduction.}

Entanglement, an intrinsically quantum effect, and its preparation are critical to envision large-scale quantum computation and simulation schemes~\cite{ncbook}. 
However, unavoidable interactions of any quantum system with its environment typically kills entanglement, thereby resulting in a mostly classical world despite fundamental principles being	inherently quantum~\cite{leggett,zurekrmp}.
The notion of ``sudden death of entanglement'' (SDE) which has been theoretically shown in~\cite{yueberly,doddhalliwell} and experimentally demonstrated subsequently~\cite{almeida} states that for typical Markovian environments, contrary to the exponential decay of one-body auto-correlations, two-body entanglement is killed completely after a finite time interval.
Despite the remarkable post-SDE progress, a robust scheme for long-distance and steady-state entanglement appears to be critical in scaling up the approaches discussed above to larger quantum networks. 
A particularly promising route is that of quantum bath engineering (QBE)~\cite{zoller96,zoller08,tomadin08,um2,girvin}.
A number of simultaneous studies focused on decoherence in non-Markovian environments \cite{mpl1,mpl2,bellomo,rlf1} showing that entanglement can be sustained over longer times through revivals and trapping.
However, these schemes allow only for transient preservation of coherence and are subject to a limited lifetime. 
A method of QBE that is versatile and effective is cavity-assisted cooling which has been investigated in the context of atomic gases~\cite{Haroche,ag1,ag2,ag3,ag4}, opto-mechanical~\cite{mec1,mec2} and spin systems~\cite{spinopt}.
Remarkable experimental progress in preparing entangled states was made with superconducting qubits~\cite{girvin2010,shankar,girvinmar,vijay,riste}, trapped ions~\cite{wineland}, macroscopic systems~\cite{krauter}, and neutral atoms~\cite{brakhane}.
There has been interesting related theoretical proposals~\cite{sorenson,chn1,chn2,spn1,plenioPRL,lamata1}. Several challenges however remain for achieving higher fidelities and scaling up these systems to many-body entangled states. 

\begin{figure}[!t]
\includegraphics[width=0.95\columnwidth]{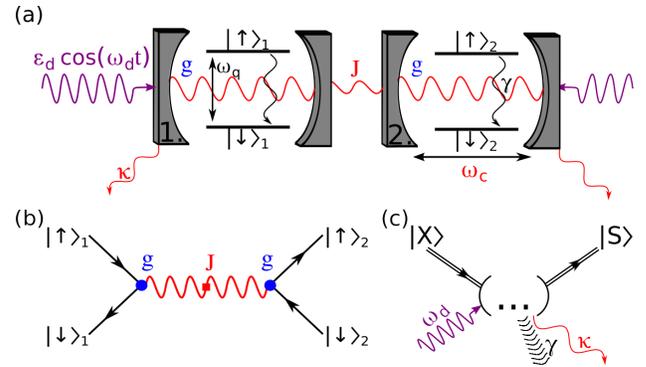}
\caption{\label{fig:diag}\footnotesize (a) Two coupled cavity-qubits driven by microwave lasers. (b) Effective qubit-qubit interaction mediated by the photons. (c) 
NESS diagram of the cooling process from any state $|X\rangle$ to the singlet $| S\rangle$. The ellipsis symbolize the transient dynamics. The by-products are heat and photon leakage.}
\end{figure}

In this work, we drive a spatially separated qubit-dimer into a target entangled two-qubit state. 
We engineer both the unitary and the dissipative dynamics \textit{via} the choice of the amplitude, phase and the frequency of coherent drive fields, all highly controllable parameters in a superconducting circuit-based platform. In particular, this involves optimally choosing a drive frequency $\omega_\rmd$ commensurate with a non-trivial emergent energy scale of the open circuit-QED dimer. We note that such a dimer setup (see Fig.~\ref{fig:diag}) has recently been experimentally realized ~\cite{dimerexp} to study a dissipation driven dynamical localization transition proposed in \cite{hetprb}. We show below that, with a suitable protocol that only involves continuous-wave drives, the system can self-stabilize to a target entangled non-equilibrium steady state, singlet or triplet, with remarkably high fidelities. 

We first introduce the driven circuit-QED dimer system ~\cite{hetprb,dimerexp} as well as the typical experimental values of various parameters. Integrating out the photonic subsystem we arrive at an effective transverse-field $XY$ model with cavity-mediated retarded interactions and whose dynamics can be described by a suitably secularized Master equation. After characterizing the emergent energetics of the system, we describe the different protocols to cool it to desired entangled states and compute the corresponding fidelities. Those can be measured \textit{via} readily available two-qubit quantum-state tomography~\cite{shankar}. We shall also discuss our findings in the broader context of many-body driven dissipative quantum systems.

\paragraph{Model.}
We consider a set of two identical two-level systems (``qubits") embedded in two identical single-mode cavities which can exchange photons, as shown in Fig.~\ref{fig:diag}. Both cavities are driven out of equilibrium by two identical coherent drives which are detuned from the cavity frequency. We consider the following Hamiltonian
\begin{align} \label{eq:H}
 H = H_\sigma + H_{\sigma,a} + H_a \;,
 \end{align}
with $H_\sigma$,  $H_{\sigma,a}$, and $H_a$ respectively the qubit, the light-matter coupling, and the photon Hamiltonians (we set $\hbar =1$)
 \begin{align}
H_\sigma =& \! \sum_{i=1}^{2}   \omega_\rmq \frac{\sigma_i^z}{2},\quad    H_{\sigma,a} =  g \! \sum_i\! \left[ a_i^\dagger \sigma^-_i + a_i \sigma^+_i  \right]  \;,
\\
   H_a =& \! \sum_i\!  \left[ \omega_\rmc a_i^\dagger a_i  +  2 \epsilon_{\rmd} \cos(\omega_\rmd t) \left( a_i + a^\dagger_i \right) \right] \nonumber \\
  &   \qquad - J \left( a_1^\dagger a_2 + a_1 a_2^\dagger  \right) \;.
  \end{align}
The two-level systems are represented by $SU(2)$ pseudo-spin operators, namely Pauli matrices, obeying $[\sigma^a_i, \sigma^b_j] = 2 \rmi \delta_{ij} \epsilon_{abc} \sigma^c_i $ where the spin indices $a,b,c = x,y,z$, the cavity indices $i,j=1,2$ and $\sigma_i^\pm \equiv (\sigma^x_i \pm \rmi\sigma_i^y)/2$. $\omega_\rmq$ is the energy splitting between ground and excited qubit states, $a^\dagger_i$ and $a_i$ are respectively the cavity-photon creation and annihilation operators, $[a_i, a_j^\dagger] = \delta_{ij}$.  $g$ is the light-matter coupling constant, $\omega_\rmc$ is the cavity frequency, and $J$ sets the strength of the photon-hopping between both cavities. 
$\epsilon_\rmd$ and $\omega_\rmd$ are respectively the strength and the frequency of both drives. 
It is convenient to work in the photon basis which diagonalizes the undriven $H_a$,
  \begin{align}
 H_{a} = \omega_\rmc^-  A^\dagger A + \omega_\rmc^+  a^\dagger a + 2\sqrt{2} \epsilon_\rmd \cos(\omega_\rmd t) (A + A^\dagger)\;,
 \end{align}
in which the symmetric (asymmetric) combination $A \equiv (a_1 +a_2)/\sqrt{2}$ [$a \equiv (a_1 -a_2)/\sqrt{2}$] corresponds to the symmetric (asymmetric) mode $\omega_\rmc^-$ ($\omega_\rmc^+$) of the coupled-cavity system with $\omega_\rmc^\pm \equiv \omega_\rmc \pm J$.
$H_{\sigma}$ alone corresponds to two independent bare qubits, the eigenstates and eigenenergies of which are given by the triplet and singlet states:
\begin{align}
\begin{array}{llll}
| T_+ \rangle  &\equiv |\uparrow \rangle_1 \otimes |\uparrow\rangle_2 \equiv |\uparrow \uparrow \rangle \;, &  E_{T_+} &= \omega_\rmq \;,  \\
| S\rangle  &\equiv \left[ |\uparrow \downarrow \rangle -  |\downarrow \uparrow \rangle \right]/\sqrt{2} \;, &  E_{S} &= 0\;, \\
| T_0\rangle  &\equiv \left[ |\uparrow \downarrow \rangle +  |\downarrow \uparrow \rangle \right]/\sqrt{2} \;, & E_{T_0} &= 0\;, \\
| T_- \rangle  &\equiv  |\downarrow \downarrow \rangle\;, & E_{T_-} &= -\omega_\rmq\;.
\end{array}
\end{align}
  
We include photon leakage outside the cavities at a rate $\kappa$. The density of states for the asymmetric and symmetric modes are $\rho_{\pm}(\omega) = -\mbox{Im } G^{\rmR}_{\pm}(\omega)/\pi$ where the retarded Green's function are given by  $G^{\rmR}_{\pm}(\omega) = 1/(\omega - \omega_\rmc^\pm + \rmi \kappa/2)$. 
We also include qubit decay and pure dephasing rates, $\gamma$ and $\gamma_\varphi$ respectively, which can be seen as the effect of coupling each qubit to a zero-temperature bosonic bath. 
These dissipative processes will play an instrumental role in our cooling scheme. 

\begin{table}
\begin{tabular}{llllll}
\hline 
$\omega_\rmc = 6$ & $\omega_\rmq = 7$ & $g = 10^{-1}$ 
$J = 10^{-1}$ &$\kappa = 10^{-4}$ &  $\gamma =  10^{-5}$ & $\gamma_\varphi = 10^{-6}$  \\ 
\hline 
\end{tabular} 
\caption{\label{tab:num}\footnotesize Typical energy scales (in $2\pi\times$GHz) that we consider.}
\end{table}

\paragraph{Effective dissipative $XY$~model.}
We treat the light-matter coupling with a second order perturbation theory in $g/\Delta$ where $\Delta \equiv \omega_\rmq - \omega_\rmc$. In Table~I, we give the typical experimental energy scales that we have in mind and which obey the hierarchy $\Delta \gg g, J \gg  \kappa, \gamma \gg \gamma_\varphi$. The pure dephasing rate is the smallest energy scale~\cite{majer}. Next, we eliminate the explicit time-dependence of the Hamiltonian by use of a rotating wave approximation. 
This can be done by first transforming the Hamiltonian in Eq.~(\ref{eq:H}) with the Schrieffer-Wolff transformation, $H \mapsto \rme^{X} H  \rme^{X^\dagger}$, where
\begin{align}
X \equiv \frac{g}{\sqrt{2}} \left[
\frac{ A(\sigma_1^+  + \sigma_2^+) }{\omega_\rmq - \omega_\rmc^-}
+
\frac{a(\sigma_1^+  - \sigma_2^+) }{\omega_\rmq - \omega_\rmc^+}
-\mathrm{h.c.}
 \right]\;.
\end{align}
Up to the order $(g/\Delta)^2$, this operation eliminates the non-linearities in $H_{\sigma,a}$ at the expense of renormalizing the qubit sector. As a consequence, the photon hopping gives rise to a cavity-mediated retarded interaction between the qubits as we shall show below [see Fig.~\ref{fig:diag}~(b)]. We then go to a rotating frame, $H \mapsto U_{\rm rot} \left[ H -\rmi \partial_t \right] U_{\rm rot}^\dagger$, with
\begin{align}
U_{\rm rot}  \equiv  \prod_{i=1}^2  \exp \left[  \rmi \omega_\rmd t \left( \frac{\sigma^z_i}{2}   + A^\dagger A  + a^\dagger a \right) \right]\;,
\end{align}
and neglect the fast rotating terms of the form $\rme^{2\rmi\omega_\rmd t} a_i^\dagger$. In the rotating frame, the bare energies are shifted by $\omega_\rmd$: $\omega_\rmq \mapsto \omega_\rmq - \omega_\rmd$, $\omega_\rmc \mapsto \omega_\rmc - \omega_\rmd$.
Noting that the drives couple to the symmetric mode of the coupled cavity system, we decompose the photon fields into mean fields plus fluctuations:
\begin{align}
A \equiv \bar A + D,\ a \equiv \bar a + d,\mbox{ and } \bar N \equiv |\bar A|^2\;,
\end{align}
where (to the lowest order in the light-matter coupling)
\begin{align}
 \bar A \simeq \frac{\sqrt{2} \, \epsilon_\rmd}{\omega_\rmd - \omega_\rmc^- + \rmi \kappa/2}  \mbox{, and } \bar a = 0\;.
\end{align}
We neglect the resulting quadratic terms in the fluctuations which couple to the qubits, \textit{e.g.} $(g/\Delta)^2 D^\dagger D \sigma_i^z$~\cite{girvin}.
More generally, we shall make sure to work in regimes for which the photon fluctuations are relatively small to comply with the validity requirements of the Schrieffer-Wolff perturbation theory~\cite{blais}.

The mean-field photonic background renormalizes the qubit sector and results in the following Hamiltonian
\begin{align} 
 \widetilde{H} =& \widetilde{H}_\sigma + \widetilde{H}_{\sigma,d} + \widetilde{H}_d\;.
 \end{align}
The qubits are now explicitly coupled, providing an experimental realization of a two-site transverse-field isotropic $XY$ model:
\begin{align}
\widetilde{H}_\sigma =  \sum_{i=1}^{2}   \boldsymbol{h} \cdot
  \frac{\boldsymbol{\sigma}_i}{2}   - \frac{1}{2} J \left(\frac{g}{\Delta} \right)^2 [\sigma_1^x \sigma_2^x + \sigma_1^y \sigma_2^y]  \;,
\end{align}
with $(h^x,h^y,h^z) \equiv ( \Omega_\rmR   , 0, \Delta_\rmq ) $
where $\Omega_\rmR \equiv 2 (g/\Delta) \epsilon_\rmd  $ and
 $\Delta_\rmq \equiv \omega_\rmq - \omega_\rmd  +  ({g}/{\Delta})^2 \left[ (\bar N 
   +1) \Delta
 + \sqrt{2} \epsilon_\rmd  \mbox{Re }\bar A \right]$. The eigenstates and eigenenergies of $\widetilde{H}_{\sigma}$ to lowest order in $g/\Delta$ are
\begin{align} \label{eq:eigen2.1}
\hspace{-2ex}
{\small
\begin{array}{llll}
| \widetilde{T}_+ \rangle  \!\!&\!\!\simeq | T_+ \rangle + \frac{\Omega_\rmR}{\sqrt{2}\Delta_\rmq} |T_0 \rangle\;, &\!\!  E_{\widetilde{T}_+}  \!\!&\!\!\simeq \Delta_\rmq + \frac{\Omega_\rmR^2}{2\Delta_\rmq}\;, \\
| \widetilde{S}\rangle \!\!&\!\!= |S\rangle \;, &\!\! E_{\widetilde{S}}  \!\!&\!\!= J (g/\Delta)^2\;,\\
| \widetilde{T}_0\rangle  \!\!&\!\!\simeq | T_0 \rangle +  \frac{ \Omega_\rmR }{\sqrt{2} \Delta_\rmq} [ |T_- \rangle  - |T_+ \rangle ] , &\!\! E_{\widetilde{T}_0}  \!\!&\!\!\simeq -J (g/\Delta)^2  \;,\\
| \widetilde{T}_- \rangle  \!\!&\!\!\simeq | T_- \rangle -  \frac{\Omega_\rmR}{\sqrt{2}\Delta_\rmq} |T_0 \rangle \;, &\!\! E_{\widetilde{T}_-} \!\!&\!\!\simeq -\Delta_\rmq - \frac{\Omega_\rmR^2}{2\Delta_\rmq} \;.
\end{array}
}
\end{align}
The degeneracy between $|S\rangle$ and $|T_0\rangle$ has been lifted by the effective qubit-qubit interaction.
The triplet states were modified by the light-matter interaction, however the eigenstate $|\widetilde{S}\rangle$ still corresponds to the singlet state of the two bare qubits. This is due by our symmetric set-up in which both qubits, cavities, and drives are identical.

Photon fluctuations on top of the coherent part couple to the qubits via
\begin{align}
\widetilde{H}_{\sigma, d} =& 
\frac{1}{2} \left(\frac{g}{\Delta}\right)^2
 \left[
\left(  \bar{A} \Delta  + \frac{\epsilon_\rmd}{\sqrt{2}}  \right) \,  D^\dagger\,
 (\sigma_1^z + \sigma_2^z) \right.
 \nonumber \\ 
&
\qquad
 + 
\left.
\left( \bar A \Delta +  \frac{\epsilon_\rmd}{\sqrt{2}} \right) d^\dagger \, (\sigma_1^z - \sigma_2^z) 
\right]  + \mathrm{h.c.}
 \label{eq:Htds}
\end{align}
which induces transitions between the eigenstates of $\widetilde{H}_\sigma$.
We treat $\widetilde{H}_{\sigma,d}$ as a perturbation to $\widetilde{H}_\sigma$ and access the dynamics of the reduced density matrix of the spin sector, $\rho_\sigma$, \textit{via} a Master equation approach. The integration over the degrees of freedom of the baths ($D$'s and $d$'s) is performed by assuming that they are in zero-temperature vacuum states. We shall see below that the deviations from this approximation can be estimated to be insignificant in the regime we study.
Importantly, as the relaxation time scale of the photon-fluctuation bath, $1/\kappa$, can be larger than the typical time scale on which $\rho_\sigma$ evolves, the dynamics of the latter is \textit{a priori} non-Markovian.
However, once a steady state is reached, $\rho_\sigma^{\mathrm{NESS}} \equiv \lim\limits_{t\to\infty} \rho_\sigma$, the usual Markovian approximation is exact~\cite{afoot,gsc,amitra} and we obtain the following non-equilibrium steady-state Master equation
\begin{align} \label{eq:Master}
\partial_t \rho_\sigma^{\mathrm{NESS}} = 0 = &
- \rmi \sum_k E_k \big{[} | k\rangle \langle k| , \rho_\sigma^{\mathrm{NESS}} \big{]}  \nonumber \\
& \qquad +
\sum_{kl} \Gamma_{k \to l} \, \mathcal{D}[|l \rangle \langle k|] \, \rho_\sigma^{\mathrm{NESS}}\;,
\end{align}
where $k$ and $l$ span the eigenstates of $\widetilde{H}_\sigma$. The Lindblad-type dissipators are defined as $\mathcal{D}[X] \rho  \equiv \left( X \rho X^\dagger - X^\dagger X \rho + \rm{h.c.} \right)/2$. The integration over the bath degrees of freedom yielded both Lamb-shift renormalizations of the energy levels (real part of self-energy) 
\begin{align} \label{eq:16}
E_k \mapsto& E_k  +  \sum_{l} \Big{[}
|\Lambda^D_{kl}|^2 \, \mbox{Re } G^{\rmR}_{-}(E_k - E_l + \omega_\rmd) \nonumber
\\
& \qquad\qquad
+
|\Lambda^d_{kl}|^2 \, \mbox{Re } G^{\rmR}_{+}(E_k - E_l + \omega_\rmd)
\Big{]}
,
\end{align}
as well as the transition rates $\Gamma_{k \to l}$ between those states (Fermi golden rule)
\begin{align}
\Gamma_{k \to  l} \equiv  \Gamma^b_{k \to  l} + \Gamma^d_{k \to  l}, \mbox{ with } 
 \Gamma^b \!= \!
{ \small
 \left( \begin{array}{cccc}
\gamma_\varphi &\gamma & \gamma & 0  \\
0 & 0 & \gamma_\varphi & \gamma \\
0 &\gamma_\varphi & 0 & \gamma \\
0 & 0 &0 & \gamma_\varphi
\end{array} \right)},
\end{align}
originating from the qubit decay and dephasing, and with the photon-fluctuation mediated part reading
\begin{align} \label{eq:15}
\Gamma_{k \to  l}^d =&
  2\pi  
   |\Lambda^D_{kl}|^2 \, \rho_{-}(E_k - E_l + \omega_\rmd) \nonumber \\
& +    2\pi 
|\Lambda^d_{kl}|^2 \, \rho_{+}(E_k - E_l + \omega_\rmd)\;,
\end{align}
where 
\begin{eqnarray}
\Lambda^D =
\lambda
{ \small
 \left( \begin{array}{cccc}
1 & \alpha & 0 & 0  \\
\alpha & 0 & 0 & \alpha \\
0 &0 & 0 & 0\\
0 & \alpha &0 & 1
\end{array} \right)}\,, \quad
\Lambda^d =  
\lambda
{ \small
 \left( \begin{array}{cccc}
0 & 0 & \alpha & 0  \\
0 & 0 & 1 & 0 \\
\alpha & 1  & 0 & \alpha \\
0 & 0 & \alpha & 0
\end{array} \right)}\,, \label{eq:matrices}
\end{eqnarray}
in which $\lambda \equiv \left({g}/{\Delta}\right)^2  \left( \bar A \Delta  + \epsilon_\rmd /\sqrt{2} \right)$
 and $\alpha \equiv \sqrt{2} \left({g}/{\Delta}\right) (\epsilon_\rmd/\Delta_\rmq) $.
Matrix rows (columns) are ordered according to $|\widetilde{T}_- \rangle$, $|\widetilde{T}_0 \rangle$, $|S \rangle$, and $|\widetilde{T}_+ \rangle$ from left to right (top to bottom).

One can re-write the Master equation (\ref{eq:Master}) as the rate equations governing the population of each of the eigenstates $|k\rangle$ of $\widetilde{H}_\sigma$, $n_k \equiv \langle k | \rho_\sigma | k\rangle$:		
\begin{align} \label{eq:pop}
 \frac{\rmd n^{\mathrm{NESS}}_k}{\rmd t} = 0 = \sum_l  n^{\mathrm{NESS}}_l \Gamma_{l\to k} -  n^{\mathrm{NESS}}_k \Gamma_{k\to l}\;,
\end{align}
which together with the conservation law $\sum_k n_k = 1$, enables us to numerically solve for the non-equilibrium steady state.

\begin{figure}[!t]
\centerline{
\includegraphics[width=0.7\columnwidth]{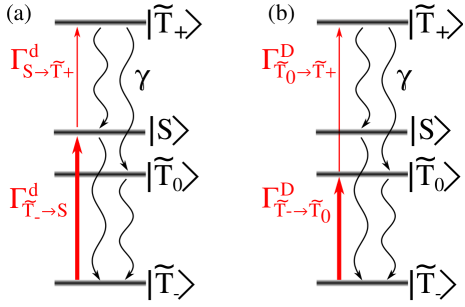}
}
\caption{\label{fig:diag2}\footnotesize (a) Mechanism for cooling to the singlet state: the photon fluctuations drive the qubit transitions from $| \widetilde{T}_- \rangle$ to $| S \rangle$ and from  $| S \rangle$ to $| \widetilde{T}_+  \rangle$ (upward arrows). Qubit decay (wavy arrows) is responsible for downward transitions and dephasing ($|\widetilde{T}_+  \rangle$) introduces transitions between  $| S \rangle$ and $| S \rangle$.
(b) Similar mechanism for cooling to $| \widetilde{T}_0  \rangle$.
}
\end{figure}

\paragraph{Cooling protocols.} 
By carefully tuning the drive frequency $\omega_\rmd$, one can engineer the photon fluctuations to trigger transitions from one renormalized eigenstate to another by Raman inelastic scattering.

We first explain the protocol to achieve convergence to the singlet state, $\rho_\sigma^{\mathrm{NESS}}  \approx |S \rangle \langle S|$, and assess its fidelity by following the population $n^{\mathrm{NESS}}_S \equiv \langle S | \rho^{\mathrm{NESS}}_\sigma | S\rangle$.
In the following discussion, we work in a parameter space for which the Lamb-shift corrections can be neglected. Nevertheless, high fidelities can also be obtained away from this regime.

By choosing $\omega_\rmd$ such that
\begin{align}
\omega_\rmd = \omega_\rmc^+ + E_{\widetilde{S}} - E_{\widetilde{T}_-}\,, \mathit{\ i.e.\ } 2\omega_\rmd \approx {\omega_\rmc^+ + \omega_\rmq} \;, \label{eq:protocol1}
\end{align}
the energy of an incoming photon can be used to add a photon in the cavity-asymmetric mode and simultaneously perform a qubit transition from $|\widetilde{T}_-\rangle$ to $|{S}\rangle$. 
Note that Eq.~(\ref{eq:protocol1}) describes a one-photon process in the rotating frame which corresponds to a two-photon process in the lab frame.
The corresponding rate is maximized, $\Gamma^d_{\widetilde{T}_- \to S} \approx 400 \, 
 (g/\Delta)^6 (\epsilon_\rmd^4/\Delta^2) /\kappa $,   and the large ratio
\begin{align}
{\Gamma_{\widetilde{T}_- \to S}}\big{/}{\Gamma_{S \to  \widetilde{T}_- }}  
\simeq {\Gamma^d_{\widetilde{T}_- \to S}}\big{/}{\gamma}  
\gg 1
\end{align}
allows for a rapid pumping from the state $| \widetilde{T}_- \rangle$ to the singlet state \textit{via} Stokes (red-shifted) scattering (see Fig.~\ref{fig:diag2}) on a very short time scale, $1/\Gamma^d_{\widetilde{T}_- \to \widetilde{S}}$, associated with the fluctuation-driven transitions.
Since the energy splitting between the states $| {S} \rangle$ and $| \widetilde{T}_+ \rangle$ is relatively close to the one between $| \widetilde{T}_- \rangle$ and $| {S} \rangle$ states, there is a simultaneous off-resonant process pumping from the state $|{S} \rangle$ to the state $| \widetilde{T}_+ \rangle$ with a rate
$\Gamma^d_{ {S} \to \widetilde{T}_+} \approx 5^2\, (g/\Delta)^2 (\epsilon_\rmd^4/\Delta^2) (\kappa/J^2)  \ll \Gamma^d_{\widetilde{T}_- \to S} $.
Both these processes are represented in Fig.~2~(a).
Note that there is no significant pumping from $| \widetilde{T}_0  \rangle$ to $| \widetilde{T}_+  \rangle$, nor from $| \widetilde{T}_-  \rangle$ to $| \widetilde{T}_0  \rangle$, because they involve producing cavity-symmetric photons which are detuned by $2J$ from the Raman resonance of Eq.~(\ref{eq:protocol1}): 
$\Gamma^D_{\widetilde{T}_- \to \widetilde{T}_0} \simeq \Gamma^D_{\widetilde{T}_0 \to \widetilde{T}_+}  \approx 100\,  (g/\Delta)^6 (\epsilon_\rmd^4/\Delta^2) (\kappa/J^2) \ll \Gamma^d_{ \widetilde{S} \to \widetilde{T}_+} \ll \Gamma^d_{\widetilde{T}_- \to S} $.
At later times, dissipation drives the system to the singlet state according to the following mechanism. 
Note first that $|\widetilde{T}_0\rangle$ is only subject to spin decay and its population decays to $0$ on a time scale set which is set by the qubit decay, \textit{i.e.} on the order of $1/\gamma$. Furthermore, the only decay channel of the singlet, from  $| S \rangle$ to $| \widetilde{T}_- \rangle$, is compensated by the rapid repumping to $| S \rangle$, see Fig.~2~(a). On the contrary, the state $| \widetilde{T}_+ \rangle$ has \emph{two} decay channels: one to $| \widetilde{T}_0 \rangle$ and another one which directly feeds  $| S \rangle$.
Altogether, the cooling to the singlet state is achieved with high fidelity and on a time scale on the order of $1/\gamma$ whenever the hierarchy 
\begin{align}
\Gamma^d_{ \widetilde{S} \to \widetilde{T}_+} \ll \gamma \ll \Gamma^d_{\widetilde{T}_- \to \widetilde{S}}
\end{align}
is obeyed.
This condition transparently elucidates that the fidelity is the result of an intricate interplay between drive, cavity decay, qubit dissipation and light-matter coupling. In particular, since $\epsilon_\rmd$ enters both sides of the hierarchy above, high fidelity is obtained in a finite window of the drive strength.

One can check \textit{a posteriori} the assumption of zero-temperature vacuum $D$ and $d$ baths by accessing the photonic properties of the system. Although the drive does not couple to the asymmetric-cavity mode, it is indirectly populated by the Raman scattering processes: one photon is produced each time a transition from $|\widetilde{T}_-\rangle$ to $|S\rangle$ or from $|S\rangle$ to $|\widetilde{T}_+\rangle$ occurs. Hence, their average number $n_d \equiv \langle d^\dagger d \rangle $ follows
\begin{align}
\frac{\rmd n_d}{ \rmd t}  = n_{T_-} \Gamma_{\widetilde{T}_-\to S} + n_{S} \Gamma_{S \to \widetilde{T}_+}  - \kappa n_d\;. \label{eq:nd}
\end{align}
Similarly for the the cavity-symmetric fluctuations, $n_D\equiv\langle D^\dagger D \rangle$, we have 
\begin{align}
\frac{\rmd n_D}{ \rmd t}  = n_{\widetilde{T}_-} \Gamma_{\widetilde{T}_-\to \widetilde{T}_0} + n_{\widetilde{T}_0} \Gamma_{\widetilde{T}_0 \to \widetilde{T}_+}  - \kappa n_D\;. \label{eq:nD}
\end{align}

\begin{figure}[!t]
\includegraphics[width=1.\columnwidth]{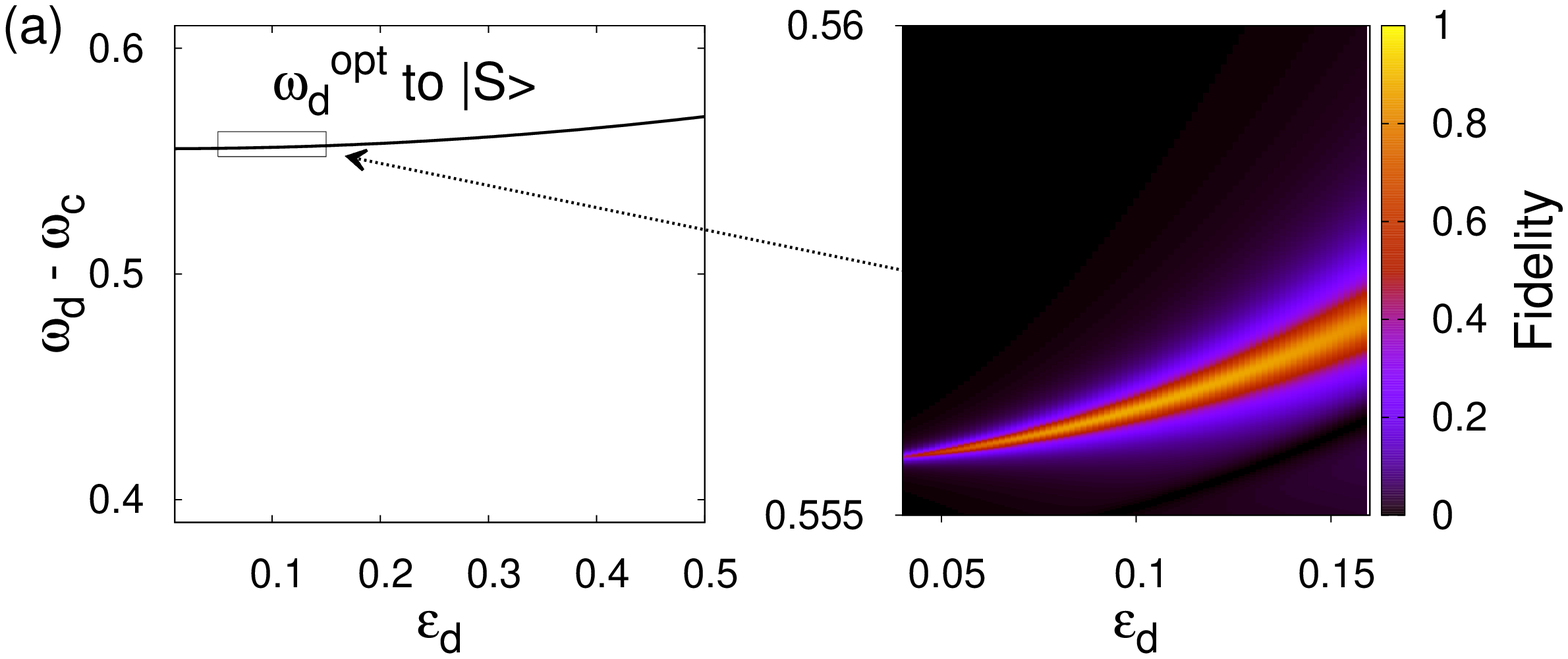}

\includegraphics[width=1.\columnwidth]{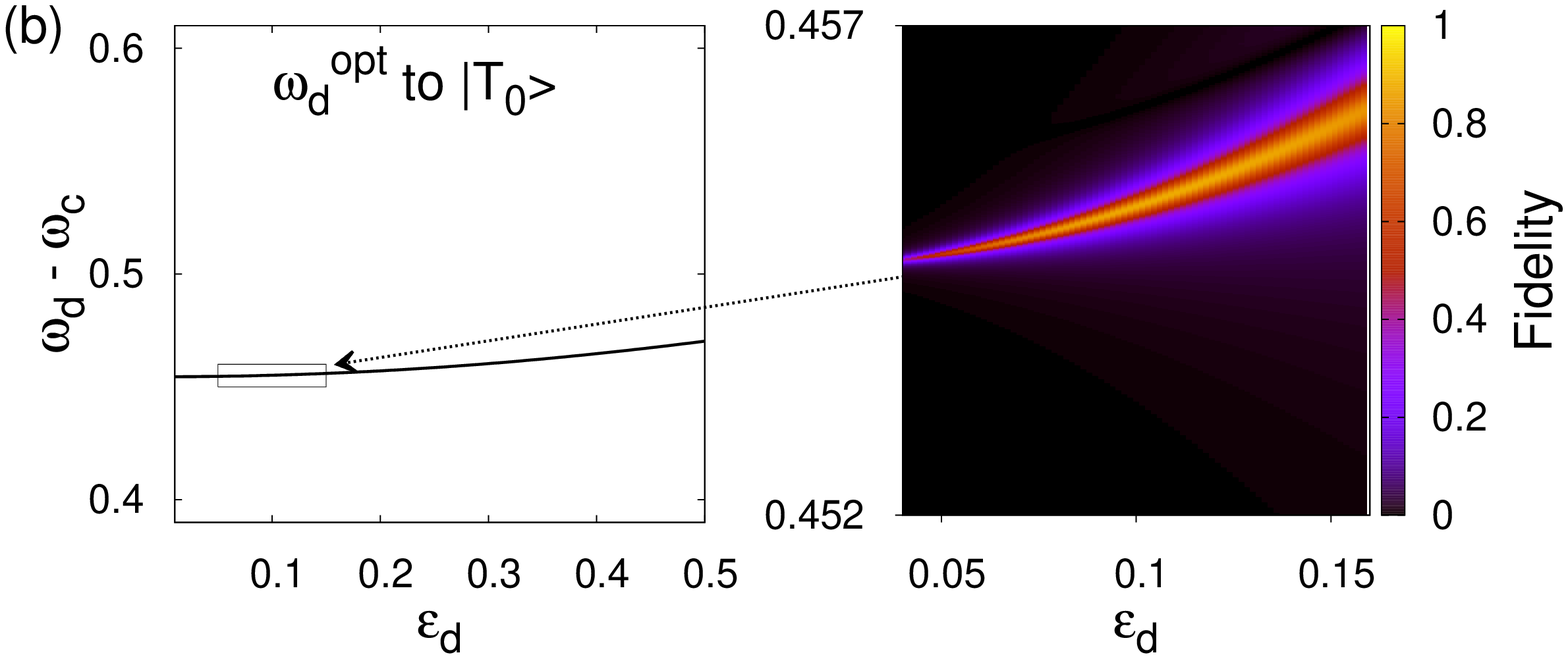}
\caption{\label{fig:tomo}\footnotesize (a) Cooling to $|S \rangle $. Left: optimal $\omega_\rmd$ solution of Eqs.~(\ref{eq:protocol1}) against the drive strength $\epsilon_\rmd$. Right: fidelity $n_S = \langle S | \rho^{\mathrm{NESS}} |S \rangle$ in a small region of $\omega_\rmd$ and $\epsilon_\rmd$. (b) Same for cooling to  $|\widetilde{T}_0 \rangle$. We used the parameters of Table~I and units are GHz.}
\end{figure}

One may wonder if driving the transition between $|\widetilde{T}_0\rangle$ and $|S\rangle$, \textit{i.e.} tuning the drive frequency $\omega_\rmd \approx \omega_\rmc + E_{\widetilde{S}} - E_{\widetilde{T}_0}$, would also result to cooling to the singlet. In fact, the lack of a mechanism to deplete the population of  $|\widetilde{T}_-\rangle$ would make it a dark state, leading to the trivial steady state $\rho^{\mathrm{NESS}} = |\widetilde{T}_-\rangle\langle \widetilde{T}_-|$.

One can also cool down the system to the other entangled eigenstate of $\widetilde{H}_\sigma$, $|\widetilde{T}_0\rangle$, by tuning $\omega_\rmd$ such that
\begin{align}
\omega_\rmd = \omega_\rmc^- + E_{\widetilde{T}_0} - E_{\widetilde{T}_-}\;.\label{eq:protocol2}
\end{align}
The cooling mechanism, Fig.~2~(b), is very similar to the one already described for the singlet and we shall not describe it in detail here. In practice, as the overlap between the original and the renormalized triplet states is $|\langle T_0  |\widetilde{T}_0  \rangle|^2 \approx 1$, this signifies that the system can be cooled to $|T_0  \rangle$ with high fidelity.

In the left panels of Fig.~3, we plot the optimal drive frequencies, solutions of Eqs.~(\ref{eq:protocol1}) and (\ref{eq:protocol2}), to cool the system to the $|S\rangle$ and $|\widetilde{T}_0\rangle$ states respectively. The difference between the optimal drive frequencies for cooling to the singlet and to the triplet state is $J$ which corresponds to the energy splitting between the two coupled-cavity modes, $(\omega_\rmc^+-\omega_\rmc^-)/2$. In the right panel, we present the non-equilibrium steady-state populations of the qubit singlet ($n_S^{\mathrm{NESS}}$) and triplet states ($n_{\widetilde{T}_0}^{\mathrm{NESS}}$)  in a small region of the drive strength $\epsilon_\rmd$ and the drive frequency $\omega_\rmd$. These quantities are readily measurable~\cite{shankar}. High fidelities can be obtained all across the parameter regime. The overall parabolic shape of the high fidelity zones stems from the quadratic dependence of the eigenenergy splittings (\textit{e.g.} $E_{\widetilde{S}} - E_{\widetilde{T}_-}$) on $\epsilon_\rmd$. 

\paragraph{Conclusions.}
The entanglement generation we described here is for qubits in two separate, tunnel-coupled cavities. Hence, in contrast to autonomous feedback schemes that typically rely on interactions in a single cavity, the qubits can in principle reside in two cavities that are connected by a waveguide, as recently demonstrated in~\cite{noda}.  The proposed scheme is therefore scalable to larger networks of spatially separated qubits and provides a general framework for the realization of synthetic non-equilibrium many-body systems subject to an engineered Liouvillian on a superconducting circuit platform~\cite{hetprb}. In particular, the model analyzed above can be extended to a dissipative transverse-field XY model on a lattice~\footnote{To be published.}. Also, identifying the signatures of entanglement in the photonic properties~\cite{knap_pra} is of high experimental importance and remains to be investigated.

We are grateful to Marco Schir\`o, Steven Girvin, Irfan Siddiqi, Shyam Shankar, Dario Gerace, Andrew Houck, Uri Vool, Mollie Schwartz, Leigh Martin and Zhen-Biao Yang for insightful discussions. This work has been supported by NSF grant DMR-115181 and The Eric and Wendy Schmidt Transformative Technology Fund.

\bibliographystyle{apsrev}
\bibliography{dimerreferences}

\begin{thebibliography}{46}
\bibitem[{afo(The coarse-graining cell in time can be made larger than
  $1/\kappa$)}]{afoot}
 (\bibinfo{year}{The coarse-graining cell in time can be made larger than
  $1/\kappa$}).
\end{thebibliography}

\end{document}